\documentclass[10pt, letterpaper, twocolumn]{article} 

\usepackage[english]{babel} 

\usepackage{microtype} 

\usepackage{amsmath,amsfonts,amsthm} 

\usepackage[svgnames]{xcolor} 

\usepackage[hang, small, labelfont=bf, up, textfont=it]{caption} 

\usepackage{booktabs} 

\usepackage{lastpage} 

\usepackage{graphicx} 

\usepackage{enumitem} 
\setlist{noitemsep} 

\usepackage{sectsty} 
\allsectionsfont{\usefont{OT1}{phv}{b}{n}} 

\usepackage{geometry} 

\usepackage[T1]{fontenc} 
\usepackage[utf8]{inputenc} 

\usepackage{XCharter} 

\usepackage{fancyhdr} 
\fancypagestyle{firstpage}{ 
	\fancyhf{}
}

\newcommand{\authorstyle}[1]{{\large\usefont{OT1}{phv}{b}{n}\color{DarkRed}#1}} 

\newcommand{\institution}[1]{{\footnotesize\usefont{OT1}{phv}{m}{sl}\color{Black}#1}} 

\usepackage{titling} 

\newcommand{\HorRule}{\color{DarkGoldenrod}\rule{\linewidth}{1pt}} 

\pretitle{
	\vspace{-30pt} 
	\HorRule\vspace{10pt} 
	\fontsize{32}{36}\usefont{OT1}{phv}{b}{n}\selectfont 
	\color{DarkRed} 
}

\posttitle{\par\vskip 15pt} 

\preauthor{} 

\postauthor{ 
	\vspace{10pt} 
	\par\HorRule 
	\vspace{20pt} 
}

\usepackage{lettrine} 
\usepackage{fix-cm}	

\newcommand{\initial}[1]{ 
	\lettrine[lines=3,findent=4pt,nindent=0pt]{
		\color{DarkGoldenrod}
		{#1}
	}{}%
}

\usepackage{xstring} 

\newcommand{\lettrineabstract}[1]{
	\StrLeft{#1}{1}[\firstletter] 
	\initial{\firstletter}\textbf{\StrGobbleLeft{#1}{1}} 
}

\usepackage[backend=bibtex,style=numeric,natbib=true]{biblatex} 

\usepackage[autostyle=true]{csquotes} 

\title{How Computer Science can Aid Forest Restoration}

\author{
	\authorstyle{G. Gordon\textsuperscript{1}, A. Holcomb\textsuperscript{2}, T. Kelly\textsuperscript{1}, S. Keshav\textsuperscript{1,3}, J. Ludlum\textsuperscript{1}, and A. Madhavapeddy\textsuperscript{1}} 
	\newline\newline 
	\textsuperscript{1}\institution{University of Cambridge}\\ 
	\textsuperscript{2}\institution{University of Waterloo}\\ 
	\textsuperscript{3}\institution{Corresponding author: email sk818@cam.ac.uk} 
}

\date{\today} 

\begin{document}

\maketitle 

\thispagestyle{firstpage} 


\lettrineabstract{The world faces two interlinked crises: climate change and loss of biodiversity. 
Forest restoration on degraded lands 
and surplus croplands can play a significant role both in sequestering carbon and 
re-establishing bio-diversity.
There is a considerable body of research and practice that addresses forest restoration. 
However, there has been little work by computer scientists to bring 
powerful computational techniques to bear on 
this important area of work,
perhaps due to a lack of awareness.
In an attempt to bridge this gap,
we present our vision of how techniques from computer science, broadly speaking, can aid current practice in forest restoration.}


\section{Introduction}
We are emerging into a post-COVID world where attention is keenly focused on the next two global crises: climate change and biodiversity loss. 
Recognizing that some critical aspects of modern society can never be fully decarbonised,
a shift to a net zero-carbon economy will necessarily require some level of carbon capture and sequestration (CCS).
Many approaches to CCS have been proposed, including pumping CO$_2$ into oil wells~\cite{vielstadte_footprint_2019}
and trapping it as chalk precipitate~\cite{matter_rapid_2016}.
These solutions, however, do not address the biodiversity crisis.
In contrast,the conservation of existing forests and the restoration of forests to degraded lands 
and surplus croplands can play a significant role both in sequestering carbon and 
re-establishing biodiversity.
In this paper, we focus only on forest restoration.

To give an idea of the scale involved, the median carbon 
sequestration potential of an agroforestry system, which combines agriculture and forestry,
is estimated to be about 95 metric tons (T) per hectare~\cite{albrecht_carbon_2003} and,
if left undisturbed, can trap this CO$_2$ for decades or even centuries~\cite{tree14}. 
This is roughly the equivalent of the emissions from 10 UK households in one year~\cite{noauthor_land_2020}.
Recent work suggests that the earth has the potential to sustain a trillion more trees planted
in approximately 900 million hectares of currently denuded land~\cite{bastin_global_2019}, 
with a carbon sequestration potential
of 205 GT, which is comparable to anthropogenic emissions of 1500 GT to date~\cite{owidco2andothergreenhousegasemissions, boden2009global}.
Thus, technological interventions to 
promote reforestation is an attractive approach to mitigate anthropogenic climate change.

In addition to carbon sequestration, native forests provide many other 
benefits both to the environment and to humans. 
Environmental benefits
include increased biodiversity,
air purification, 
flood prevention, 
and slope stabilization. 
Human benefits include
jobs in forestry and agroforestry, 
improving public health, providing food, 
and providing a source of sustainable building materials. 
For these reasons, reforestation has become a topic of great current interest.
For example, the World Economic Forum announced
in January 2020 an initiative to plant 1 trillion trees~\cite{seddon2021getting}.

Unfortunately, reforestation is not simply a matter of 
planting saplings or seeds in the ground and walking away~\cite{castroprecision, lamb_large-scale_2014, seddon2021getting}. 
\begin{itemize}
  
    \item Reforestation guided purely by economic criteria  results
    in monoculture plantations that can neither sequester carbon for 
    the long term nor support biodiversity~\cite{lamb_large-scale_2014, lewis_restoring_2019}. They also
    lack resilience to disease, pests, and climate change. 
    
    \item Successful tree-planting, but of fast-growing species such as pine or 
    eucalypts, can result in not only draining of the water table~\cite{lu_ecological_2018} (and depletion of soil nutrients in the case of eucalypts) 
    but also increased fire risk~\cite{natgeofire}.

    \item A top-down technocratic approach that ignores the land's owners and other local actors 
    (who may even have been responsible for 
    deforestation in the first place) is frequently doomed to failure~\cite{davis_ecologies_2018}. 

    \item Even multi-species, long-term reforestation motivated by climate credits may encourage fraud due to lack of trustworthy monitoring \cite{song2019even}.  
    
    \item There is ongoing debate about the accuracy of the models used to project the CCS benefits from forests~\cite{yanai2020improving}.
\end{itemize}

Fortunately,
over the past two decades, practitioners in the discipline 
of Forest and Landscape Restoration (FLR) have developed a deeper understanding of 
reforestation approaches that
balance the needs of the environment and local actors \cite{lamb_large-scale_2014, chazdon2017policy, brancalion2017beyond, mansourian2019putting, castroprecision}.
At the risk of oversimplification, this work can be summarized as follows.

First, for ecological restoration and carbon sequestration, as opposed to mere tree harvesting,
it is important to combine appropriate policies (such as a guarantee of long-term land tenure)
with institutional frameworks (such as participatory management of communal forests).
Importantly, reforestation
outcomes must be economically sustainable for all stakeholders.
To ensure this,
local actors must be engaged participants in the planning process.

Second, there is a continuing need for the improved monitoring of 
trees, other vegetation, soil, and wood products
at all spatial scales. This monitoring can not only help identify forests under
threat of degradation and deforestation but also help evaluate forest health, raise the alarm for interventions and provide
a basis for reliable administration of carbon credit payments. 

These two areas are precisely where computer science, broadly speaking, can help FLR practitioners, especially 
when there is a need for very large scale and resources are limited. 
In the remainder of this paper,
we present an outline of the some potential areas for computer scientists to be 
involved in forest restoration
(Section \ref{sec:lifecycle}), along with a handful of near-term
research opportunities (Section \ref{sec:opp}) to demonstrate the great potential for work in this area.
Note that some interventions can reuse existing computational tools, while others will require fundamentally new techniques. 

\section{FLR lifecycle}
\label{sec:lifecycle}
\begin{figure}
  \includegraphics[width=0.6\columnwidth]{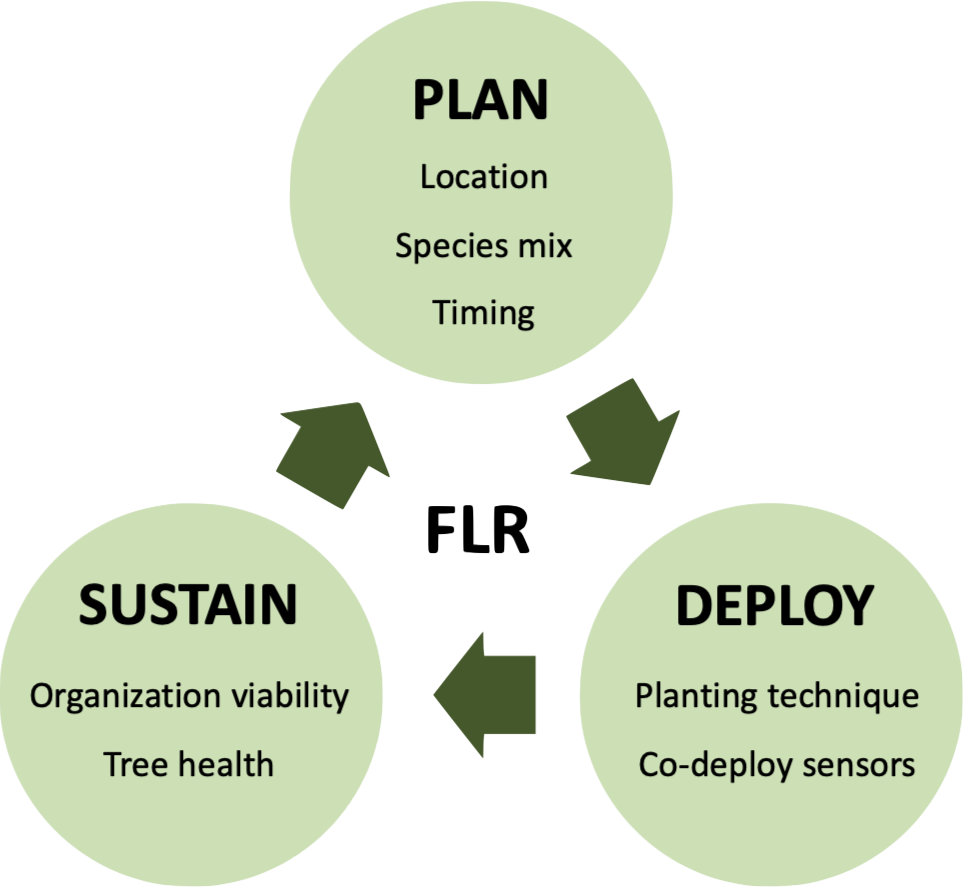}
 \caption{The Forest and Landscape Restoration lifecycle.}
\end{figure}
 \label{fig:lifecycle}
 
A forest restoration project goes through three distinct phases~\cite{castroprecision} (Figure \ref{fig:lifecycle}).
In the \textit{planning} phase (\S\ref{sec:planning}), critical decisions need to be made about where 
to plant trees (\textit{location}) and what species to plant (\textit{species mix})
so as to restore woodlands in a way that best mimics pre-existing ecosystems. 
In the \textit{deployment} phase (\S\ref{sec:deployment}) trees are planted, potentially along with ground-based monitoring equipment such as sensors. 
Finally, in the \textit{sustenance} phase (\S\ref{sec:sustenance}), the health of both the trees themselves
as well as the FLR organization need to maintained.

We discuss these phases in more detail next,
outlining how computer science can help in each phase,
and subsequently
cover research opportunities  (\S\ref{sec:opp}).

\subsection{Planning}
\label{sec:planning}
In the planning phase, forestry experts and local actors collaborate to plan the tree locations, species mix, and time of planting. 


\textbf{Choice of location}: 
Trees would ideally be planted to rehabilitate degraded lands 
that are otherwise unsuitable for agriculture or human habitation, making best use of existing
rootstock and undisturbed pockets of mature forest that can provide animal, plant, and fungal colonizers~\cite{lamb_large-scale_2014}. 
At a macroscopic level, AI-based analysis of satellite data and LiDAR 
can be used to create 
3D elevation maps to identify potential areas of forest degradation~\cite{yao_tree_2012,andersen_estimating_2005}. 

Once suitable land has been identified, the local climate and micro-climate must be considered in choosing exact planting locations. 
Significant factors include the land slope, levels of rainfall and groundwater flow, wind speed and flow, different types of soil and different native species of vegetation~\cite{gunter_determinants_2009}. 
Given the complexity, participation by local actors can help 
avert poor decisions (Section \ref{sec:vis}).

At a microscopic level, careful consideration of the existing vegetation and soil state is required, since disturbing soil actually 
releases existing stores of greenhouse gases, as well as increasing the risk of erosion, particularly if 
protective soil crusts are destroyed~\cite{ladron_de_guevara_simulated_2014}.
Moreover, forests that soak up rainwater for their own growth may result in shriking water
supplies downstream.
Hence, mathematical optimization can be used to choose locations that balance the cost of
breaking ground and loss of water supply with long-term benefits from planting in a particular location. 
Note that successful FLR projects need to consider the human element of land function, 
social participation, and natural resources management~\cite{fao_review_2004,barr_political_2012}. 
Optimization models must, therefore, include the 
socio-environmental factors attached to the physical locations, 
such as the type of land tenure, the current functions of the land and potential impacts associated with the land in question.

\textbf{Choice of species mix:}
Poor choice of trees can further damage the environment~\cite{cao_damage_2010}.
Moreover, 
a diverse population of trees is important for the health of a forest; even a 10\% loss of 
diversity can lead to a 3\% loss of productivity~\cite{liang_positive_2016}, as well as a higher risk of disease~\cite{liu_protecting_2003}.  
A productive species mixture emerges if there is complementary resource usage among different trees~\cite{kelty_role_2006}, and is influenced by the micro-climate~\cite{baker_microclimate_2014} and expected future climate~\cite{noce_likelihood_2017}. Species diversity complicates the planning process due to the need for more sophisticated analysis of mapping data~\cite{donoghue_remote_2007}.
There is some evidence that the correlation between biodiversity above 
ground and soil carbon sequestration rates is dependent on the landscape and climate context~\cite{pichancourt_growing_2014}. 
Hence, there is a need for accurate modelling of multi-species forests (discussed in \S\ref{sec:agents}).



\subsection{Deployment}
\label{sec:deployment}

In this phase, both trees and monitoring equipment must be deployed safely across the restoration area.

\textbf{Planting:} Trees can be planted as saplings,
seeds, or pelletised seeds, i.e., seeds embedded in a pellet that contains soil, fertilizer, anti-grazing chemicals, and sometimes a dye for visibility~\cite{lamb_large-scale_2014}.
Planting can be done by humans, or by helicopters and planes.
In many cases, planting must be preceded by burning weed cover or scarifying the earth using tractors or other equipment.

Given the labour intensive nature of planting, it is natural to consider
the use of automation.
Handling, storage and delivery practices can affect seedling establishment, and environmental factors and the planting microsite directly affect the physiological response of saplings immediately after planting~\cite{grossnickle_bareroot_2016}. 
If the species is one that can be grown from seed, drone fleets can be used
to seed large areas rapidly \cite{noauthor_dendra_nodate}. 
It is conceivable that a form of robotics may be used to plant either seeds or saplings, as well as for weeding in the first few years\footnote{E.g. The Milrem tree planter from Estonia: https://milremrobotics.com/product/robotic-forester-planter/}; 
this is a potential topic for future research. 

\textbf{Sensor co-deployment:} In addition to planting the trees, sensor technology should be simultaneously deployed. 
Remote sensing technologies such as RFID and LoRA can be used to gather microscopic level data such as soil salinity, which can aid the identification of vulnerable areas or vegetation that requires immediate intervention but also play a vital role in ongoing data collection, feedback and extrapolation. 
We discuss the types of sensors that might be deployed in this forested environment later (\S\ref{sec:sensors}).

\subsection{Sustenance}
\label{sec:sustenance}
While many socio-economic and socio-environmental factors determine the long-term
success or failure of an FLR project
~\cite{le_what_2014,le_more_2012},
it is clear that planted forests need to be monitored for several decades to ensure
long-term carbon sequestration. 
Organizations responsible for forest management also need to be financially sustainable.

\textbf{Forest sustainability:}
Once seeds or saplings are established and thriving, we need to ensure their continued health for maximum carbon storage, as well as broader markers such as biodiversity. 
Environmental success indicators include: a suitably high survival rate;
the structure of the planted vegetation (canopy, ground, litter and shrub cover); species diversity; and ecosystem functions (overall soil health, quantities of groundwater and of course, carbon sequestration)~\cite{le_what_2014}. All of these indicators need to be appropriately monitored  (\S\ref{sec:sensors}).

\textbf{Organization viability:}
Organizations responsible for forest management need to be financially sustainable and free of fraud. 
Creating a system of trusted proofs for carbon certification may be a viable
source of funding for the long term. Local sensor data and global satellite sweeps
combined with digitally verified contracts and transactions could go some way to 
eliminate the fraud present in existing carbon certification programs~\cite{frunza_fraud_2013}.  
We discuss some options later (\S\ref{sec:blockchain}).

\section{Research opportunities}
\label{sec:opp}

Computer Science has an important role to play in scaling FLR programs to the global scale required to plan, deploy and sustain a trillion trees over the coming decades. We now discuss several research opportunities based on current understanding of FLR (\S\ref{sec:lifecycle}).

\subsection{Trustable carbon sequestration credits}
\label{sec:blockchain}

Forest-based carbon sequestration has recently started to attract substantial funding. 
For example, 
the UK Forestry Commission announced the Woodland Carbon Guarantee~\cite{wcg} in 2019, a £50m scheme to provide landowners with the option to sell sequestered CO2 from newly planted trees back to the government for a guaranteed price for the next 30 years. As schemes such as this grow worldwide, the opportunity for fraud has also followed~\cite{frunza_fraud_2013}.  
Hence, there is an opportunity to use distributed ledger technology to inject trust into the system \cite{tapscott2017blockchain}.  
Automation via trustworthy distributed ledgers~\cite{androulaki_hyperledger_2018,allombert_introduction_2019} would ensure that such schemes are economically sustainable over the decades needed to realise the carbon sequestration potential of newly planted forests.

A distributed ledger for carbon sequestration credits would need to register ownership of a
a reforestation project and then reliably track its carbon sequestration over time. 
The project owner would be permitted to 
generate annual sequestration certificates for the prior year of forest regrowth; 
if the project were to fail, then revenues from the sale of these certificates would
dry up, incentivizing the appropriate behaviour~\cite{tree14}.
Satellite observations could be used to reliably observe
project outcomes.
Moreover, cloud-based data archiving, curation, and analysis of satellite observations  would provide 
ongoing detailed information to certificate purchasers about the progress of the scheme. 
The design of such a global, scalable, and trustworthy system is an open area for research.

\subsection{Visualization and collaboration technology}
\label{sec:vis}

It is critical to the long-term success of an FLR project that local actors and policy makers are 
happy with the eventual outcomes.
FLR practitioners must also elicit knowledge of local conditions and constraints from local stakeholders, including indigenous communities,
during the planning phase ~\cite{galuemalemana_assessment_2015}.
These two objectives can be met by presenting
\textit{visualizations} of potential outcomes and using \textit{collaboration technology}
to allow planning participants to contribute their knowledge.

While technologies for collaboration are well known~\cite{schauer2010reviewing},
there are research opportunities in developing innovative visualization technology 
specifically for FLR.
For example, state-of-the-art visualization using Augmented Reality might be beneficial in this context~\cite{west_metatree_2012}.  
We believe that AR can help in creating common ground and make the outcomes more real to participants, policy makers, and stakeholders. Once local stakeholders are involved in the process, it is natural to crowdsource data from them to scale our data collection.  Mobile apps that encourage citizen science 
(such as Seek by iNaturalist and WWF~\cite{seekapp}) could provide ongoing crowdsourced data feeds to augment 
the planning and sustenance processes, as well as educate participants about the progress and benefits they are 
realizing as the newly planted forests grow.


Visualizations can also be used to present large amounts of
data that has been gathered from satellite- and ground-based observations.
One useful direction may be
for computer scientists to construct an online knowledge elicitation, communication, and collaboration platform to involve all the stakeholders and work across disciplines~\cite{lazoschavero_stakeholders_2016}.




\subsection{Monitoring forests}
\label{sec:sensors}


Monitoring forests using sensor systems is critical from many perspectives: deciding on interventions, measuring forest health,
and validating carbon credits.
\textit{Macroscopic} monitoring from satellites in well known, and more recently
\textit{mesoscopic} monitoring, such as LiDAR scans from planes or drones, 
provide an excellent source of data for recognising individual trees \cite{lee_graph_2017}, which is vital for trustable sequestration credits and estimating current and potential carbon sequestration.  
Further data can be
gathered at the \textit{microscopic} scale from on-the-ground sensors 
helping us to determine the health of individual trees (e.g. identifying disease, insect or animal damage, or failure to thrive) and enabling rapid intervention if required.
As trees grow, the micro-climate changes, strongly affecting the plant communities \cite{baker_microclimate_2014}, therefore ongoing measurements of temperature, humidity and other environmental parameters are also needed~\cite{zellweger_advances_2019}.  

However, many sensors in use today in agricultural contexts are not suitable for long-term mass-scale forest deployment.
Forests environments are less controlled and consistent than agricultural ones, and access to deployed sensors is comparatively limited. Thus, the sensors need to be self-managing, robust and long lived. 
Sensors may make different trade-offs between energy usage and lifespan. For example, active medium-lifespan sensors (e.g. spectroscopy for measuring soil quality) used during sapling establishment could be deployed alongside passive long-lived sensors designed to last for the expected lifetime of the trees (e.g. temperature and humidity via chipless RFID \cite{noauthor_case_2016}).
Additional sensors may also be considered outside of those normally associated with agriculture, for example to account for dead organic matter, estimate carbon content of soil and litter, and to monitor for poaching, theft, vandalism and possible fraud.
Thus, there is ample scope for research into appropriate sensor system design.


Another direction of sensor system design is for monitoring, reporting, and verification (MRV) of forest-based carbon projects.
The existing protocol to measure forest carbon sequestration, for example in the UK Woodland Carbon Code (WCC)~\cite{noauthor_uk_nodate}, 
is almost entirely manual, with a suggested material list that includes a smartphone, GPS or compass, printed paper spreadsheets, and measuring tape. The process is time-intensive and may be quite costly, with surveyors recommended to begin their work 6-12 months before the scheduled verification deadline. Moreover, the UK is not alone: both a 2016 review of research into REDD+ programs and a 2018 report on forest-based carbon projects highlighted MRV as a key challenge \cite{mbatu_redd_2016,grimault_key_2018}. 

In the short term, camera-mounted, GPS-enabled drones with a connection to the cloud, working alongside human surveyors, might offer a dramatic improvement for data collection, management, and analysis. Longer term, these labelled data sets could further the development of machine learning and vision algorithms to measure tree growth or infer species and health. Eventually, this data might be integrated with existing satellite or LiDAR imagery, and could be collected completely autonomously.


\subsection{Agent-based models for forest simulation}
\label{sec:agents}

An agent-based model (ABM) can simulate the growth of a multi-species
forest from seed to maturity, taking into account the local soil condition, environmental variability, and inter-species interaction~\cite{marechaux2017individual, parry2012large}.
It can play a key role throughout the lifecycle of tree restoration projects,
from planning to sustenance. In the planning phase (\S\ref{sec:planning}), 
an ABM allows competing forest restoration approaches to be evaluated 
ahead of planting. It may help determine what species mix to plant in 
a given area or which land area is most likely to be successful in reforestation, 
incorporating concerns spanning multiple disciplines, such as the level of stakeholder 
engagement and the soil characteristics. Robustness stress testing could also be conducted
for fire risks and potential climate changes. 
Such a simulation tool can also be used to help sustain a forest project 
(\S\ref{sec:sustenance}), for example by allowing foresters to simulate interventions for disease management or understand how a forest will respond to extreme weather events. 
It can also play a key role in the financial element of project sustenance. 
Many carbon credit programs require demonstrating increased sequestration against a hypothetical baseline~\cite{noauthor_methodology_2013}, 
which can be more easily and reliably established through trusted, peer-reviewed models.

There are existing agent-based modelling packages that allow large-scale forest simulation, for example TROLL~\cite{marechaux2017individual}, but new CS techniques will be needed to 
scale up the simulations and utilize cloud-scale clusters.
An ideal ABM would also allow large scale evaluation of emergent system behaviour 
while automatically integrating macroscopic (satellite), mesoscopic (LiDAR) and microscopic 
(soil condition) data to calibrate and verify the simulations. 
For example, satellite images could be used to identify species types in existing forest and re-forested areas. 
These areas could have their growth from saplings simulated, using our knowledge of growth processes~\cite{lam_simulating_2005,de_reffye_model_1995,soderbergh_algorithms_2003}, 
while controlling error propagation by feeding in high-res satellite imagery~\cite{rejou-mechain_upscaling_2019}. 
By adjusting the model's parameters, the simulation should show the same crown cover (for some metric of similarity) as reality. This combination of growth processes with multiple data scales would allow new reforestation techniques to be accurately developed.


\section{Conclusions}
The role of reforestation in carbon sequestration is well recognized.
A naive approach to reforestation, however, can lead to numerous undesirable consequences,
which can be averted by relying on proven methodologies from the discipline of Forest and Landscape Restoration.
We have identified several broad areas where computer science tools and techniques can help FLR practitioners.
We also identify CS research problems in the specific areas of forest simulation, trustable carbon credits, visualization, and monitoring. We hope that 
the new tools and techniques created from this research
will give FLR practitioners new degrees of freedom 
in restoring degraded lands and combating climate change. 
We conclude by quoting the UK’s Committee on Climate Change, which 
in its recent assessment criterion for new potential land-use policies asks
``\textit{Is the [reforestation] policy feasible to implement, monitor and verify over time?}''~\cite{noauthor_land_2020}. By extending the 
scope of what is feasible, computer scientists can aid in developing
effective reforestation policies not just
in the UK but around the world.

\printbibliography[title={References}]
\end{document}